\documentclass[a4paper,11pt,twoside]{article}

\usepackage{geometry} 
\usepackage{amsmath}
\usepackage{amssymb}
\usepackage{graphicx}
\usepackage{epstopdf}
\usepackage{color}

\makeatletter
\renewcommand{\fnum@table}{\textbf{\tablename~\thetable}}
\renewcommand{\fnum@figure}{\textbf{\figurename~\thefigure}}
\makeatother

\newlength{\myem}
\newcommand{\sep}[1]{#1}
\newcounter{mysubequation}[equation]
\renewcommand{\themysubequation}{\alph{mysubequation}}
\newcommand{\mytag}{\stepcounter{mysubequation}%
\tag{\theequation\protect\sep{\themysubequation}}}
\newcommand{\globallabel}[1]{\refstepcounter{equation}\label{#1}}

\makeatletter
\renewcommand{\section}{\@startsection{section}{1}{0em}%
        {-3.5ex \@plus -1ex \@minus -.2ex}% 
        {2.3ex \@plus.2ex}%
        {\normalfont\large\bfseries}}
\renewcommand{\subsection}{\@startsection{subsection}{2}{0em}%
        {-3.25ex\@plus -1ex \@minus -.2ex}%
        {1.5ex \@plus .2ex}%
        {\normalfont\bfseries}}
\renewcommand{\subsubsection}%
        {\@startsection{subsubsection}{3}{0em}%
        {-3.25ex\@plus -1ex \@minus -.2ex}%
        {1.5ex \@plus .2ex}%
        {\normalfont\itshape}}
\makeatother

\newcommand{\TeV}{\,\mathrm{TeV}}
\newcommand{\GeV}{\,\mathrm{GeV}}
\newcommand{\MeV}{\,\mathrm{MeV}}

\newcommand{\ecm}{e\,\mathrm{cm}}
\newcommand{\Fig}[1]{Fig.~\ref{fig:#1}}

\newcommand{\Figs}[1]{Figs.~\ref{fig:#1}}

\newcommand{\Eq}[1]{Eq.~(\ref{eq:#1})}
\newcommand{\eq}[1]{eq.~(\ref{eq:#1})}

\newcommand{\eqs}[1]{eqs.~(\ref{eq:#1})}
\newcommand{\vev}[1]{\left\langle #1\right\rangle}
\newcommand{\ord}[1]{\mathcal{O}\left( #1 \right)}

\newcommand{\interskip}{\medskip}
\DeclareMathOperator{\im}{Im}

\DeclareMathOperator{\diag}{Diag}

\newcommand{\tilgu}{{\tilde g}_u}
\newcommand{\tilgd}{{\tilde g}_d}
\newcommand{\tilgup}{{\tilde g}^\prime_u}
\newcommand{\tilgdp}{{\tilde g}^\prime_d}
\newcommand{\tilgus}{{\tilde g}^*_u}
\newcommand{\tilgds}{{\tilde g}^*_d}
\newcommand{\tilgups}{{\tilde g}^{\prime *}_u}
\newcommand{\tilgdps}{{\tilde g}^{\prime *}_d}

\newcommand{\SISSA}{SISSA/ISAS and INFN, I--34013 Trieste, Italy}
\newcommand{\CERN}{CERN, Department of Physics, Theory Division \\
  CH--1211 Geneva 23, Switzerland}

\newcommand{\preprintdate}{October 2005}
\newcommand{\preprintnumber}{
CERN--TH/2005--196 \\
SISSA 74/2005/EP
}
 
\newcommand{\titletext}{Electric Dipole Moments in Split Supersymmetry} 
\newcommand{\authortext}{\large G.F. Giudice$^{\, a}$ and A. Romanino$^{\, b}$
\medskip\\\em\normalsize 
$\mbox{}^a$ \CERNÄ
\\[0.1\baselineskip] 
$\mbox{}^b$ \SISSA}
\newcommand{\abstracttext}{We perform a quantitative study of the
neutron and electron electric dipole moments (EDM) in Supersymmetry,
in the limit of heavy scalars. The leading contributions arise at two
loops. We give the complete analytic result, including a new
contribution associated with $Z$--Higgs exchange, which plays an
important and often leading role in the neutron EDM.  The predictions
for the EDM are typically within the sensitivities of the next
generation experiments. We also analyse the correlation between the
electron and neutron EDM, which provides a robust test of Split
Supersymmetry.}

\title{
\normalsize
\begin{tabular}[t]{l}%\hepnumber\\
\preprintdate\end{tabular}
\hspace*{\fill}
\begin{tabular}[t]{l}\preprintnumber\end{tabular}
\vspace{3\baselineskip}\\\Large\bfseries\titletext\bigskip}
\author{\begin{minipage}[t]{0.8\textwidth}
\normalsize\centering\authortext
\end{minipage}}
\date{}

\begin{document}

\bigskip
\maketitle
\begin{abstract}\normalsize\noindent
\abstracttext
\end{abstract}\normalsize\vspace{\baselineskip}

% \clearpage

% \noindent

\section{Introduction}

The electric dipole moments (EDMs) of the Standard Model (SM) fermions
are powerful probes of physics beyond the SM. Once the strong CP
problem has been taken care of, the SM predictions for the EDMs
of quarks and leptons are at least 7 orders of magnitudes
below~\cite{Pospelov:05} the present experimental
limits~\cite{Harris:99a,Romalis:00a,Regan:02a}.  The situation is
drastically different in supersymmetric extensions of the SM. The
supersymmetry-breaking terms involve many new sources of
CP-violation. Particularly worrisome are the phases associated, in the
universal and
flavour-diagonal case, to the invariants ${\rm arg}(A^* M_{\tilde g})$ and
${\rm arg}(A^* B)$. Such
phases survive in the universal limit in which all the flavour
structure originates from the SM Yukawas. If these phases are of order
one, the electron and neutron EDMs induced at one-loop by
gaugino-sfermion exchange are typically (barring accidental
cancellations~\cite{Brhlik:98a}) a couple of orders of magnitude above
the limits~\cite{Abel:01a,Demir:03a,OPRS}, a difficulty which is
known as the supersymmetric CP problem.

The Split limit
of the MSSM~\cite{AD,Giudice:04a,Arkani-Hamed:04a} does not present
a supersymmetric CP problem.
Heavy sfermions suppress the dangerous one-loop contributions to a
negligible level. Nevertheless, some phases survive below the
sfermion mass scale and, if they do not vanish for an accidental or a
symmetry reason, they give rise to EDMs that are
safely below the experimental limits, but sizeable enough to be well
within the sensitivity of the next generation of
experiments~\cite{Arkani-Hamed:04a}. Such contributions only arise at
the two-loop level, since the new phases appear in the
gaugino-Higgsino sector, which is not directly coupled to the SM
fermions. 

In this paper, we perform a quantitative study of the neutron and
electron EDMs in the limit of Split Supersymmetry. First, we compute 
the different
contributions to the light quark and electron EDMs,
the only relevant CP-violating operators. Indeed, quark chromoelectric
dipoles and the gluon Weinberg operator~\cite{Weinberg} cannot be
generated at two loops. For the EDM,
the original CP-violation in the gaugino-Higgsino sector is
communicated to the SM fermions by gauge boson and Higgs exchanges,
specifically by i) $\gamma h$ , ii) $WW$, or iii) $Zh$ exchange. No
other possibilities are allowed at the two-loop level.

The $\gamma h$ exchange has been widely studied in the literature in
several contexts~\cite{Barr:90a,Chang:98a,Chang:02a,Pilaftsis:02a}.
The case of Split Supersymmetry was considered
in ref.~\cite{Arkani-Hamed:04a}. The $WW$ exchange has also been
studied in different
limits~\cite{Marciano:86a,Kadoyoshi:96a,Deshpande:05a}. An exact
2-loop computation has been performed in the context of Split
Supersymmetry in ref.~\cite{Chang:05a} (see also 
ref.~\cite{Lopez-Mobilia:94a} for a computation in the context of a 
two-Higgs doublet model). Our results, for which we give 
explicit analytic expressions,
differ from those in ref.~\cite{Chang:05a}. 
Moreover, we identify a third, important contribution due to $Zh$
exchange. The $Zh$ contribution is suppressed in the case
of the electron EDM by a $1-4\sin^2\theta_W$ factor, but it plays an
important role in the neutron EDM. In fact, the $Z h$ contribution is
always comparable and often larger than the $\gamma h$ one (which in
turn is tipically larger than the $WW$ contribution). We have also
recomputed the QCD renormalization effect, correcting a mistake present
in the previous literature.

\section{General expressions for the EDMs}

CP-violating phases can enter the effective Lagrangian below the sfermion mass 
scale $\tilde
m$ through the Yukawa couplings (which are irrelevant for our study),
the $\mu$-parameter, the gaugino masses $M_i$,
$i=1,2,3$, or the Higgs-Higgsino-gaugino couplings $\tilgu$,
$\tilgd$, $\tilgup$, $\tilgdp$ in 
\begin{equation}
  \label{eq:couplings}
  -\mathcal{L} = 
  \sqrt{2}\left( \tilgu H^\dagger \tilde W^a T_a \tilde H_u +  \tilgup
  Y_{H_u}H^\dagger \tilde B \tilde H_u +  \tilgd H_c^\dagger
  \tilde W^a T_a \tilde H_d +  \tilgdp 
  Y_{H_d}H_c^\dagger \tilde B \tilde H_d \right) +{\rm h.c.},
\end{equation}
where $H_c = i\sigma_2 H^*$, $T_a$ are the $SU(2)$ generators, and
$Y_{H_u}=-Y_{H_d}=1/2$. 
% We
%will consider a phase convention in which the
%$\mu$-term and the gaugino masses are real and positive. 
The Higgs vev is in its usual form $\vev{H} = (0,v)^T$, 
with $v\sim 174\GeV$. The gaugino and Higgsino mass parameters $M_{1,2}$ and $\mu$, and
the couplings $\tilgu$, $\tilgd$, $\tilgup$, $\tilgdp$ are in general complex.
However, only three phases are independent and
they are associated to the invariants $\phi_1 = \arg(\tilgups \tilgdps
M_1\mu)$,
$\phi_2 =
\arg(\tilgus \tilgds M_2\mu)$, $\xi
= \arg(\tilgu \tilgds \tilgdp \tilgups )$. The tree-level matching
with the full theory above $\tilde m$ gives $\arg(\tilgu) = \arg(\tilgup)$,
$\arg(\tilgd) = \arg(\tilgdp)$, and therefore $\xi = 0$, thus leaving only
two independent phases. Moreover, in
most models of supersymmetry breaking the phases of $M_1$ and $M_2$ are
equal, in which case there is actually only one CP-invariant.

\interskip

In terms of mass eigenstates, the relevant interactions are 
\begin{multline}
  \label{eq:interactions}
  -\mathcal{L} = \frac{g}{c_W} \overline{\chi^+_i} \gamma^\mu
  (G^R_{ij}P_R + G^L_{ij}P_L)\chi^+_j Z_\mu \\
  + \left[ g \overline{\chi^+_i} \gamma^\mu (C^R_{ij}P_R +
    C^L_{ij}P_L) \chi^0_j W^+_\mu +\frac{g}{\sqrt{2}} \overline{\chi^+_i}
    (D^R_{ij}P_R +D^L_{ij}P_L) \chi^+_j h +\text{h.c.}  \right] ,
\end{multline}
where
\globallabel{eq:mixings}
\begin{align}
  &G^L_{ij} = V^{\phantom{\dagger}}_{iW^+}c_{W^+}V^\dagger_{W^+j} +
  V^{\phantom{\dagger}}_{ih_u^+}c_{h_u^+}V^\dagger_{h_u^+j} &
  -&{G^{R}_{ij}}^* = U^{\phantom{\dagger}}_{iW^-}c_{W^-}U^\dagger_{W^-j} +
  U^{\phantom{\dagger}}_{ih^-_d}c_{h_d^-}U^\dagger_{h_d^-j} \mytag \\
  &C^L_{ij} = -V^{\phantom{*}}_{iW^+} N^*_{jW_3} +\frac{1}{\sqrt{2}}
  V^{\phantom{*}}_{ih^+_u}N^*_{jh^0_u} &
  &C^R_{ij} = -U^*_{iW^-} N^{\phantom{*}}_{jW_3} -\frac{1}{\sqrt{2}}
  U^*_{ih^-_d}N^{\phantom{*}}_{jh^0_d} \mytag \\
  g&D^R_{ij} = \tilgu^* V_{ih^+_u}U_{jW^-} +\tilgd^*
  V_{iW^+}U_{jh^-_d} & &D^L=(D^R)^\dagger . \mytag
\end{align}
In eq.~(\ref{eq:mixings}a), $c_f = T_{3f} -s_W^2 Q_f$ ($s_W^2\equiv
\sin^2\theta_W$) is
the neutral current coupling coefficient of the fermion $\tilde
f$ and, accordingly, $c_{W^\pm}=\pm \cos^2\theta_W$, 
$c_{h_u^+,h_d^-}= \pm(1/2-s^2_W)$.
The matrices $U$, $V$, $N$ diagonalize the complex chargino and neutralino
mass matrices, $M_{+} = U^TM^D_{+} V$, $M_0 = N^T N^D_0 N$, where
$M^D_+ = \diag(M^+_1,M^+_2)\geq 0$, $M^D_0 =
\diag(M^0_1,\ldots,M^0_4)\geq 0$ and 
\begin{equation} 
  \label{eq:charginos}
  M_+ =
  \begin{pmatrix}
    M_2 & \tilgu v \\
    \tilgd v & \mu
  \end{pmatrix}, \quad M_0 =
  \begin{pmatrix}
    M_1 & 0 & -\tilgdp v/\sqrt{2} & \tilgup v/\sqrt{2} \\
    0 & M_2 & \tilgd v/\sqrt{2} & -\tilgu v/\sqrt{2} \\
    -\tilgdp v/\sqrt{2} & \tilgd v/\sqrt{2} & 0 & -\mu \\
    \tilgup v/\sqrt{2} & -\tilgu v/\sqrt{2} & -\mu & 0
  \end{pmatrix}.
\end{equation}

\interskip

\begin{figure}
\begin{center}
\includegraphics[width=\textwidth]{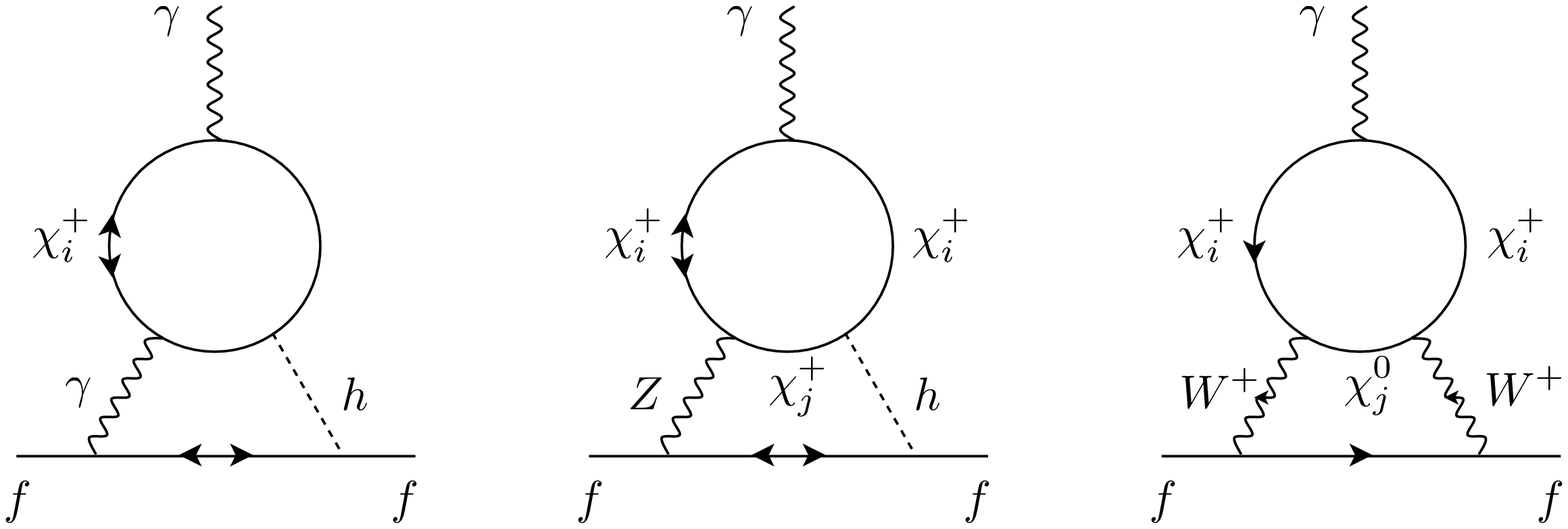}
\end{center}
\caption{Two loop contributions to the light SM fermion EDMs. The third diagram is for a down-type fermion $f$.}
\label{fig:feynman}
\end{figure}

In Split Supersymmetry, fermion EDMs are generated only at two loops, 
since charginos and neutralinos, which carry the information of CP violation,
are only coupled to gauge and Higgs bosons. 
To identify all possible diagrams contributing to the EDM, 
let us first consider the
case in which $M_{1,2},\mu \gg M_W$. After we integrate out charginos
and neutralinos at one-loop, we generate some effective couplings among
SM bosons. These can be described in terms of gauge-invariant, CP-violating
operators. There are 5 dimension-6 such operators: $\epsilon_{abc}
{\widetilde W}^a_{\mu\nu}W^{b\nu\rho}W^{c\mu}_{\rho}$, $H^\dagger H
{\widetilde W}^a_{\mu\nu}W^{a\mu\nu}$,
$H^\dagger H
{\widetilde B}_{\mu\nu}B^{\mu\nu}$,
$D_\mu H^\dagger D_\nu H{\widetilde B}
^{\mu\nu}$,
$D_\mu H^\dagger T_a D_\nu H{\widetilde W}^{a\mu\nu}$, where $W^a_{\mu\nu}$
and $B_{\mu\nu}$ are the $SU(2)$ and $U(1)$ gauge strengths, and
${\widetilde W}^a_{\mu\nu}$ and ${\widetilde B}_{\mu\nu}$ are their duals.
The effective couplings relevant to generate sizable two-loop contributions to the
EDM must contain 3 fields, with at least one photon and at most one Higgs
boson. The previously-listed operators induce only the effective couplings
$\gamma\gamma h$, $\gamma Z h$, and $\gamma WW$. Notice that
CP-violating couplings of the kind $\gamma \gamma \gamma$,   
$\gamma \gamma Z$ and $\gamma ZZ$ are not generated (in particular,
the CP-violating operator $B_{\mu \nu}B^{\nu \rho}B^\mu_\rho$ identically
vanishes unless there are three different abelian gauge fields). The absence
of these couplings is also confirmed by an explicit one-loop calculation.
Once we insert the effective couplings in a loop, we obtain 3 different
diagrams contributing at the two-loop level to the EDM
of the light SM fermion $f$, shown in \Fig{feynman}. We therefore have
\begin{equation}
  \label{eq:d}
  d_f = d^{\gamma H}_f + d^{ZH}_f + d^{WW}_f ,
\end{equation}
\globallabel{eq:dipoles}
\begin{align}
  d^{\gamma H}_f & =  \frac{e Q_f\alpha^2}{4\sqrt{2}\pi^2s^2_W}
  \im(D^R_{ii}) \frac{m_f M^+_i}{M_W m^2_H} f_{\gamma H}(r^+_{iH}) \mytag \\
  d^{ZH}_f &=  \frac{e \left(T_{3f_L} -2 s^2_W Q_f \right)
\alpha^2}{16\sqrt{2}\pi^2c^2_Ws^4_W}  \im \left( D^R_{ij}G^R_{ji}
    -D^L_{ij}G^L_{ji} \right) \frac{m_f M^+_i}{M_W m^2_H}
  f_{ZH}(r^{\phantom{\dagger}}_{ZH},r^+_{iH},r^+_{jH}) \mytag \\
  d^{WW}_f & = \frac{eT_{3f_L} \alpha^2}{8\pi^2s^4_W} \im \left(
    C^L_{ij}C^{R*}_{ij} \right) \frac{m_f M^+_i M^0_j}{M^4_W}
  f_{WW}(r^+_{iW},r^0_{jW}) , \mytag
\end{align}
In \eqs{dipoles}
the sum over the indexes $i,j$ is understood, $Q_f$
is the charge of the fermion $f$, $T_{3f_L}$ is the third component of
the weak isospin of its left-handed component. Also,
$r_{ZH} = (M_Z/m_H)^2$, $r^+_{iH} = (M^+_i/m_H)^2$,  $r^+_{iW} = 
(M^+_i/M_W)^2$, $r^0_{iW} = (M^0_i/M_W)^2$, where $m_H$ is the Higgs mass,
and the loop functions are given by
\globallabel{eq:loopfunctions}
\begin{align}
  f_{\gamma H}(r) &= \int^1_0\frac{dx}{1-x}\,
  j\left(0,\frac{r}{x(1-x)}\right) \mytag \\
  f_{Z H}(r, r_1,r_2) &= \frac{1}{2}\int^1_0 \frac{dx}{x(1-x)}\,
  j\left(r,
    \frac{x r_1+(1-x) r_2}{x(1-x)}\right) \mytag \\
  f_{WW}(r_1,r_2) &= \int^1_0\frac{dx}{1-x}\, j\left(0,
    \frac{x r_1 + (1-x) r_2}{x(1-x)}\right) . \mytag
\end{align}
The symmetric loop
function $j(r,s)$ is defined recursively by
\begin{equation}
  \label{eq:basicfunctions}
  j(r) = \frac{r\log r}{r-1}, \quad j(r,s) = \frac{j(r)-j(s)}{r-s}.
\end{equation}
In the determination of $f_{ZH}$ we have used the symmetry of
$\im(D^R_{ij}G^R_{ji})M^+_i$ and $\im(D^L_{ij}G^L_{ji})M^+_i$ under
$i\leftrightarrow j$. \Eq{dipoles} differs from the result in 
ref.~\cite{Chang:05a}. Analytic expressions for the functions in 
\eqs{loopfunctions} are given in the appendix.

The parameters entering the expression of the quark EDM in \eqs{dipoles}
have to be evaluated at the chargino mass scale $M^+$. The renormalization to 
the scale $\mu$ 
at which we evaluate the neutron EDM matrix element is determined
by the anomalous dimension of the operator $\bar q \sigma_{\mu \nu}
\gamma_5 q F^{\mu\nu}$. We find
\begin{equation}
  \label{eq:eta}
d_q(\mu)=\eta_{\text{QCD}}~ d_q(M^+),\quad
\eta_{\text{QCD}}=\left[ \frac{\alpha_s(M^+)}{\alpha_s(\mu)}\right]^{\gamma /2b},
\end{equation}
where the $\beta$-function coefficient is $b=11-2n_q/3$ and $n_q$ is the number
of effective light quarks. The anomalous-dimension coefficient is
$\gamma =8/3$. To eliminate the quark mass dependence in the short-distance 
contribution, it may be preferable to consider the ratio
$d_q/m_q$. Its renormalization is given by 
\begin{equation}
  \label{eq:eta2}
\frac{d_q}{m_q}(\mu)=\eta_{\text{QCD}}^{\gamma^\prime /\gamma}\,
\frac{d_q}{m_q}(M^+),
\end{equation}
with  $\gamma^\prime =32/3$. For $\alpha_s (M_Z)=0.118\pm 0.004$ and
$\mu=1\GeV$ (the scale of the neutron mass), we find 
$\eta_{\text{QCD}}= 0.75$ for $M^+ =1\TeV$ and $\eta_{\text{QCD}}= 0.77$ for 
$M^+ =200\GeV$. The error in $\alpha_s (M_Z)$ gives an uncertainty
on $\eta_{\text{QCD}}$ of about 2\%, while we expect an uncertainty of
about 5\% from next-to-leading order effects. 
Notice that the value of $\eta_{\text{QCD}}$ obtained here is different than 
what computed in ref.~\cite{nano} and generally used in the literature. 
Indeed, ref.~\cite{nano} incorrectly uses the opposite sign for $\gamma$.
Our result gives a QCD renormalization coefficient about a factor of 2 smaller
than usually considered, and it agrees with the recent findings of
ref.~\cite{degrassi}.

To express the neutron EDM in terms of the quark EDMs, we use the
results of QCD sum-rule techniques~\cite{Pospelov:99a,Pospelov:00a}:
\begin{equation}
  \label{eq:neutron}
  d_n = (1\pm 0.5)\left[\frac{f_\pi^2 m_\pi^2}{(m_u+m_d) (225\MeV)^3}
  \right] \left(\frac{4}{3} d_d -\frac{1}{3} d_u \right) ,
\end{equation}
where $f_\pi \approx 92\MeV$ and we have neglected the 
contribution of the quark chromoelectric
dipoles, which does not arise at the two-loop level in the heavy-squark
mass limit. Note that $d_n$ depends on the light quark masses only through
the ratio $m_u/m_d$, for which we take the value $m_u/m_d = 0.553\pm
0.043$. 

\section{Expansions in the heavy-chargino limit}

We now discuss the result in the limit in which the $R$-symmetry
breaking scale, determining gaugino and Higgsino masses, is
larger than $M_Z$ and $m_H$. A leading-order perturbative expansion
of \eq{dipoles} in powers of $|M_{1,2}\mu|/M_Z^2$ and $|M_{1,2}\mu|/m_H^2$
(keeping all orders in $|M_{1,2}/\mu|$ and in $M_Z/m_H$) gives
\globallabel{eq:expansion}
\begin{align}
  d^{\gamma H}_f & \simeq \frac{eQ_f\alpha m_f}{8\pi^3}
   \frac{\tilgu \tilgd}{ M_2 \mu}  \sin \phi_2 F_{\gamma H}\left( \frac{M^2_2}{\mu^2},\frac{M_2\mu}{m_H^2}\right) \mytag \\
  d^{Z H}_f & \simeq \frac{e\left(T_{3f_L} -2s^2_WQ_f  \right)
   \alpha m_f}{16\pi^3
    c^2_W }\frac{\tilgu \tilgd}{ M_2 \mu} \sin \phi_2 
    F_{Z H} \left(\frac{M_Z^2}{m_H^2}, \frac{M^2_2}{\mu^2},\frac{M_2\mu}{m_H^2}
\right) \mytag \\
  d^{WW}_f & \simeq \frac{e T_{3f_L} \alpha m_f}{16\pi^3
    s^2_W } \left[ \frac{\tilgu \tilgd}{ M_2 \mu} \sin \phi_2
    F^{(2)}_{WW}\left( \frac{M^2_2}{\mu^2},\frac{M_2\mu}{m_H^2}\right) + \frac{\tilgup \tilgdp} {M_1 \mu} \sin \phi_1
    F^{(1)}_{WW}\left( \frac{M^2_1}{\mu^2},\frac{M_1\mu}{m_H^2}\right) \right]
  , \mytag
\end{align}
where ${\tilde g}_{u,d}$, ${\tilde g}_{u,d}^\prime$, $M_{1,2}$ and $\mu$ now
indicate the absolute value of the corresponding
quantity and the functions $F_{\gamma H}$, $F_{ZH}$, $F_{WW}$ are
given in the appendix. As long as $M_1/M_2<1$ (as, for instance, in 
the case of gaugino masses
unifying at the GUT scale), the second term in eq.~(\ref{eq:expansion}c)
is suppressed with respect to the first term and numerically it is not
very significant.
Notice that eqs.~(\ref{eq:expansion}) explicitly exhibit
the dependence on the two CP-violating invariants $|\tilgu \tilgd / M_2 \mu|
\sin \phi_2$ and $|\tilgup \tilgdp / M_1 \mu|
\sin \phi_1$. Because of the suppression of the second term in 
eq.~(\ref{eq:expansion}c), both the electron and neutron EDM are mostly
characterized by a single invariant.

While eqs.~(\ref{eq:expansion}a,b) can be obtained from an expansion
at the first order in $v/M_{1,2}$, eq.~(\ref{eq:expansion}c) arises only at
the second order. This is because the origin of $d^{WW}_f$ can be
traced back to the vertices $\tilde W^{\mu\nu} D_\mu H D_\nu H$,
$\tilde B^{\mu\nu} D_\mu H D_\nu H$ in the unbroken electroweak symmetry
phase, requiring two insertions of the Higgs vev. 
The additional $M/v$ factor in eq.~(\ref{eq:dipoles}c) leads to
a contribution to the EDM which is parametrically of the same order 
of $d^{\gamma H}_f$ and $d^{ZH}_f$ in the $v/M$ expansion. Notice also that
the coefficients of the potentially large logarithms of $|M_{1,2}\mu| 
/m_H^2$ correspond to the anomalous dimensions that mix the EDM operator
with the CP-violating dimension-6 operators obtained from integrating out the
supersymmetric particles well above the weak scale. 

The relative importance of the three contributions to $d_f$ in \eq{d}
can be estimated from \eqs{expansion}. Let us consider for definitess
the case $M_2=\mu$. By keeping only terms enhanced by a large
$\log(M_2\mu/m^2_H)$ in the expressions~(\ref{eq:Fexpansion}) for
$F_{\gamma H}$, $F_{ZH}$, $F^{(2)}_{WW}$, we obtain $F_{ZH} \approx
F_{\gamma H}(-s_W^2 +3/4)$ and $F^{(2)}_{WW} \approx -F_{\gamma
H}/4$. As a consequence, we find
\begin{equation}
\label{eq:largelog2}
\frac{d_f^{ZH}}{d_f^{\gamma H}} \approx \frac{(T_{3f_L}-2s^2_WQ_f)(3-4s_W^2)}
{8c^2_WQ_f}   \quad \frac{d_f^{WW}}{d_f^{\gamma H}} \approx
-\frac{T_{3f_L}}{8s^2_WQ_f}  \quad \text{($M_2=\mu$)},
\end{equation}
where in the expression  for $d_f^{WW}$ we have neglected the subleading
second term in
eq.~(\ref{eq:expansion}c).
Numerically, \eq{largelog2} gives $d_e^{ZH} \approx 0.05\,d_e^{\gamma
H}$, $d_e^{WW} \approx -0.3 \,d_e^{\gamma H}$ and $d_n^{ZH} \approx
d_n^{\gamma H}$, $d_n^{WW} \approx -0.7 \,d_n^{\gamma H}$. These
simple estimates show the importance of the $ZH$ contribution to the
neutron EDM. A detailed numerical analysis of the relative
importance of the different contributions to the electron and neutron
EDMs is given in the next Section. The qualitative estimates above are
in a remarkably good agreement in the large $M_2=\mu$ limit.

\section{Numerical analysis}

We now perform a numerical analysis of the full results for the EDM in
\eqs{dipoles}. We consider a standard unified framework for the
gaugino masses at the GUT scale. By using the RGEs given 
in refs.~\cite{Giudice:04a,Arvanitaki:04a}, the parameters in \eqs{dipoles}
can be expressed in terms of the single phase $\phi\equiv\phi_2$ and
the four following positive parameters: $M_2$, $\mu$ (evaluated at
the low-energy scale), $\tan\beta$, and
the sfermion mass scale $\tilde m$. 
In first approximation, the dipoles depend on
$\beta$ and $\phi$ through an overall factor $\sin2\beta\sin\phi$.
Therefore, in order to maximize the effect,
we choose $\tan\beta = 1$ and the phase $\phi$ evaluated at the low-energy 
scale such that $\sin\phi = 1$. Notice that if the ratio $|\mu /M_2|$
is much larger or much smaller than one, the renormalization effects 
which mixes $\mu$ and $M_2$ (peculiar of Split Supersymmetry) tend to
suppress the effective phase. In this case, a maximal CP violation can only 
be achieved with very particular choices of the initial values for the higgsino
and gaugino masses. Once we have fixed $\sin 2\beta \sin \phi =1$,
we are then left with
the three dimensionful parameters $M_2$, $\mu$,
$\tilde m$. The overall sfermion scale $\tilde m$ enters only
logarithmically through the RGE equations for ${\tilde g}_{u,d}$,
${\tilde g}_{u,d}^\prime$. We choose to present the
results as contour plots in the $M_2$--$\mu$ plane and set $\tilde m =
10^9\GeV$, which is consistent with the cosmological bounds given in 
ref.~\cite{Pierce}. 

\begin{figure}
\begin{center}
\includegraphics[width=0.8\textwidth]{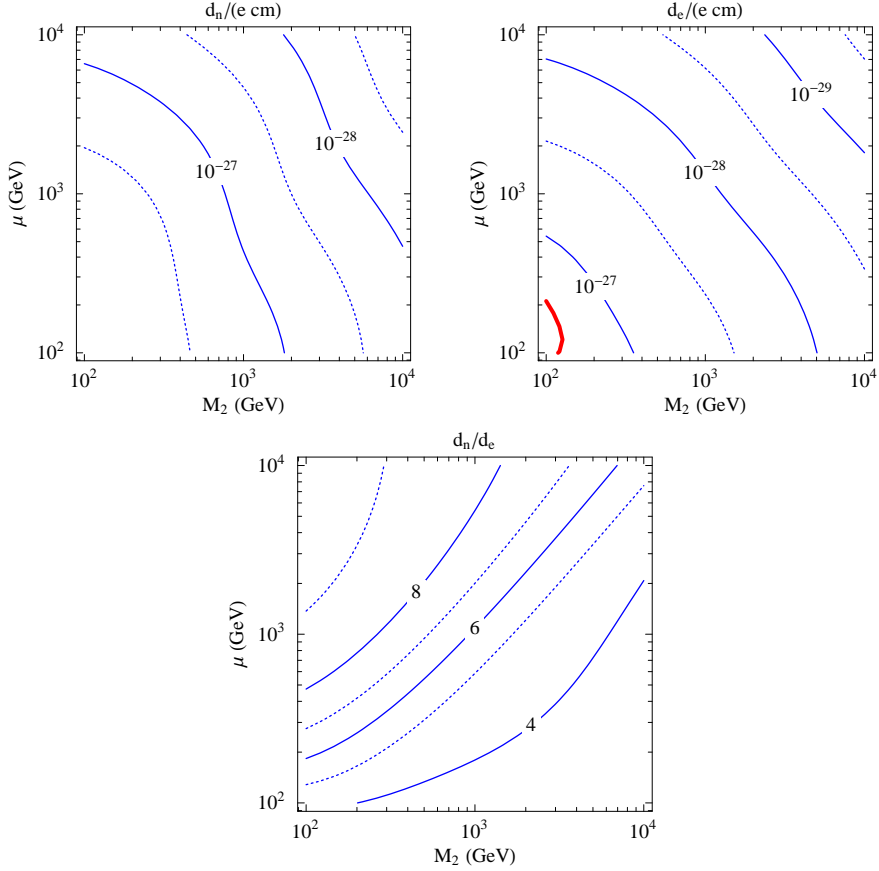}
\end{center}
\caption{Prediction for $d_n$, $d_e$, and their ratio $d_n/d_e$. We
have chosen $\tan\beta = 1$, $\sin\phi = 1$, and $\tilde m =
10^9\GeV$. The results for $d_n$ and $d_e$ scale approximately linearly with 
$\sin 2\beta \sin\phi$, while the ratio is fairly independent of $\tan\beta$,
$\sin\phi$ and $\tilde m$.
The red thick line corresponds to the present experimental
limit $d_e< 1.6\times 10^{-27}\ecm$~\cite{Regan:02a}.}
\label{fig:edm}
\end{figure}

\Fig{edm} shows the prediction for the electron EDM, the neutron EDM,
and their ratio $d_n/d_e$.  The red thick line corresponds to the
present experimental limits $d_e< 1.6\times
10^{-27}\ecm$~\cite{Regan:02a}, while the limit $d_n<0.63\times
10^{-27}\ecm$~\cite{Harris:99a} does not pose a constraint on the
parameters shown in \Fig{edm}. In the Split limit, and assuming
gaugino mass unification, all EDMs are controlled by a single
phase. The results for $d_e$ and $d_n$ shown in \Fig{edm} scale approximately
linearly with $\sin 2\beta \sin \phi$.

A robust test of Split Supersymmetry can be performed if both the 
electron and the neutron EDM are measured. Indeed,
in the ratio $d_n/d_e$ the dependence on $\sin\phi$,
$\tan\beta$ and $\tilde m$ approximately cancels out.
This can be easily understood from eqs.~(\ref{eq:expansion}) which
show that, as long as the chargino and neutralino masses are sufficiently
larger than $M_Z$, the only dependence of $d_n/d_e$ on ${\tilde g}_{u,d}$,
${\tilde g}_{u,d}^\prime$, and $\sin \phi_{1,2}$ comes from the
existence of the $M_1$-dependent term in $d_f^{WW}$, see 
eq.~(\ref{eq:expansion}c). This term, as previously discussed, is numerically
small. 
Nevertheless, because of the different loop functions associated to
the different contributions, the ratio $d_n/d_e$ varies by a
$\ord{100\%}$ factor when the $M_2$ and $\mu$ are varied in the range
spanned in the Figures. Still, the variation of $d_n/d_e$ 
is comparable with the
theoretical uncertainty on the determination of $d_n$ in terms of
quark EDMs in \eq{neutron} due to the hadronic matrix element, and is
significantly smaller than the variation in the ordinary MSSM prediction,
even in the case of universal phases (see
e.g.~ref.~\cite{AL}\footnote{Note that the $ZH$ contribution is missing in
the analysis of the Split Supersymmetry case in ref.~\cite{AL}, which leads
to a stronger correlation between $d_e$ and $d_n$.}). On the other
hand, the usual tight correlation between the
electron and muon EDMs, $d_\mu/d_e = m_\mu/m_e$ persists.

\begin{figure}
\begin{center}
\includegraphics[width=0.8\textwidth]{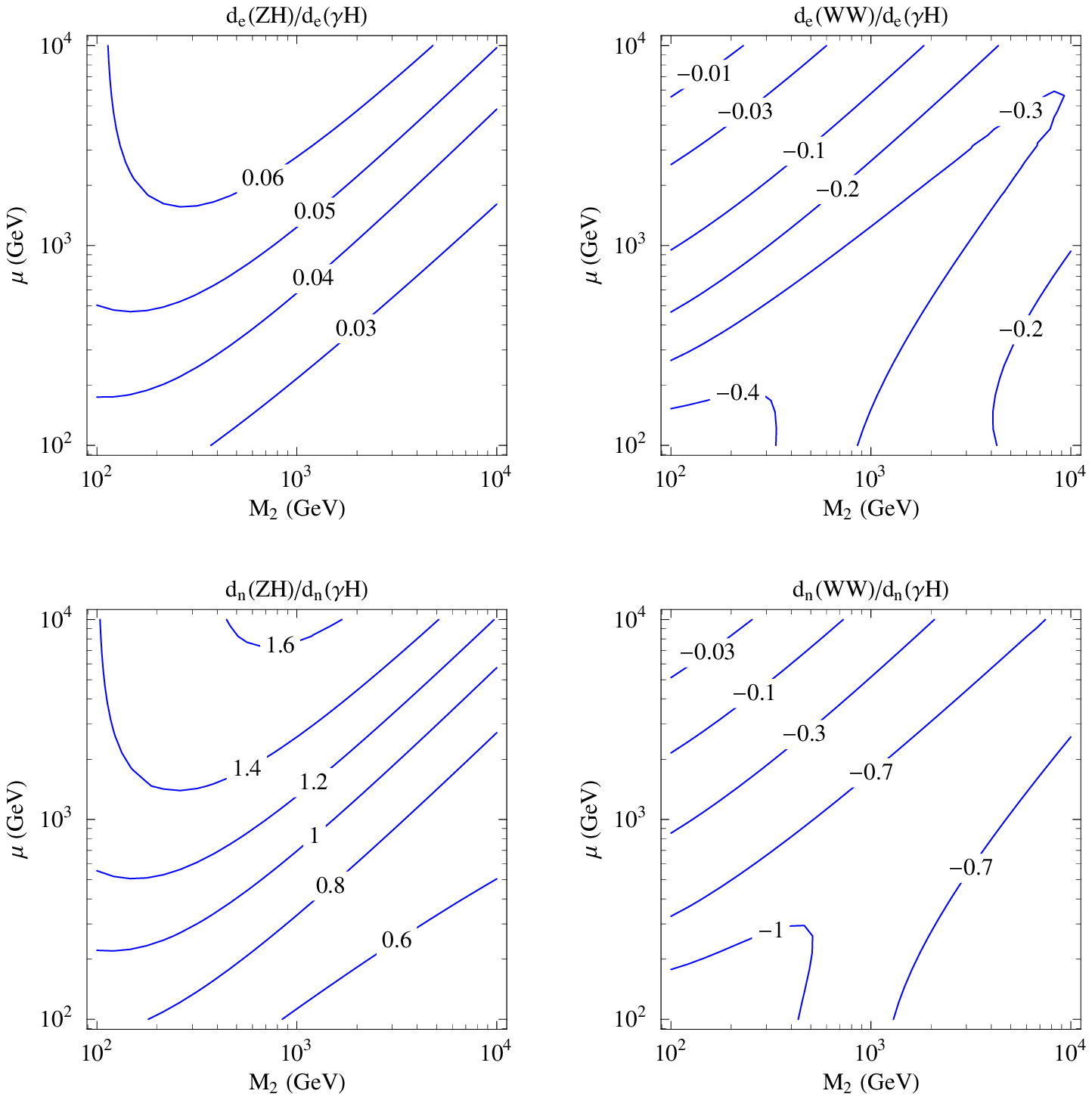}
\end{center}
\caption{Relative importance of the different contributions to the
EDMs. We have chosen $\tan\beta = 1$, $\sin\phi = 1$ and $\tilde m =
10^9\GeV$, but the result depends only very weakly on this choice.}
\label{fig:con}
\end{figure}

\Fig{con} shows the relative importance of the different contributions
to the EDMs. For the same reason explained above, the results shown in
\Fig{con} are fairly independent of $\sin\phi$,
$\tan\beta$ and $\tilde m$. 
As anticipated, the $ZH$ contribution to the electron EDM
is suppressed by the $T_{3f_L}-2s^2_WQ_f$ factor. On the other hand,
the corresponding contribution to $d_n$ is always important and
represents the largest contribution in a significant portion of the
parameter space shown in \Fig{con}. The $WW$ contribution is also
sizable, especially in the case of the neutron EDM, but is typically
smaller than the $ZH$ or $\gamma H$ ones. While the $ZH$ and $\gamma
H$ contributions always add constructively, the $WW$ contribution has
an opposite sign. However, also due to the $ZH$ contribution, its 
size is not large enough to flip the
sign of the overall electron or neutron EDM. 

\section{Conclusions}

In summary, we performed a quantitative study of the neutron and
electron EDMs in Split Supersymmetry. Clearly, our results also apply
to the MSSM in any limit in which the contributions involving sfermion
exchange are suppressed. Our result for the $WW$ exchange differs from
the one in the literature. Moreover, we find a new contribution
associated to $Zh$ exchange, which plays an important and often
leading role in the neutron EDM. We have also given the correct value
for the QCD renormalization of the quark EDM.
We performed an analytical and
numerical analysis of our results, summarized in \Figs{edm}
and~\ref{fig:con}. The correlation between the electron and neutron
EDMs is found to be stronger than in standard supersymmetric
scenarios, and it may become a crucial experimental test for
Split Supersymmetry. Still, we find an $\ord{100\%}$ variation of $d_n/d_e$ in
the parameter space we considered, which is comparable with the
hadronic uncertainty on the determination of $d_n$ in terms of quark
EDMs. The results summarized in \Fig{edm} are quite promising in the
light of the expected impressive improvement of the experimental
sensitivities in the years to come and represent one of the most
relevant windows on Split Supersymmetry.

\section*{Acknowledgments}

We thank A. Ritz for useful discussions and the authors
of ref.~\cite{Chang:05a} for communications. A.R. wishes to thank
the CERN Theory Division for the kind hospitality during the first
stage of this work.

\section*{Appendix}

We write below the analytic expression of the loop functions in \eqs{dipoles} and~(\ref{eq:loopfunctions}).
\begin{equation}
f_{\gamma H}(r) =\frac{1}{\sqrt{1-4 r}}\left[\log r \; \log
\frac{\sqrt{1-4 r}-1}{\sqrt{1-4 r}+1} 
+\text{Li}_2\left(\frac{2}{1- \sqrt{1-4r}}
\right)
-\text{Li}_2\left(\frac{2}{1+ \sqrt{1-4r}}\right) \right]
\end{equation}
\begin{multline}
f_{ZH}(r,r_1,r_2) =\frac{1}{r-1}\left\{  
g(r,r_1,r_2)-g(1,r_1,r_2) \right. \\
\left. -\frac{\log r}{2(x_1-x_2)} \left[\log\left(1-\frac{1}{x_1}\right) -\log\left(1-\frac{1}{x_2}\right)\right]\right\}
\end{multline}
\begin{multline}
f_{WW}(r_1,r_2) = \frac{1}{\hat y_1 -\hat y_2}\left\{
(r_2-\hat y_1) \left[ \log r_2 \log \left(1-\frac{r_2}{\hat y_1}\right)
- \log r_1 \log \left(1-\frac{r_1}{\hat y_1}\right)
\right. \right.\\
\left. +
\text{Li}_2\left( \frac{r_2}{\hat y_1}\right)-
\text{Li}_2\left( \frac{r_1}{\hat y_1}\right) \right] + 
(r_2-\hat y_2) \left[ \log r_1 \log \left(1-\frac{r_1}{\hat y_2}\right) \right. \\
\left.\left. - \log r_2 \log \left(1-\frac{r_2}{\hat y_2}\right) +
\text{Li}_2\left( \frac{r_1}{\hat y_2}\right)-
\text{Li}_2\left( \frac{r_2}{\hat y_2}\right) \right] \right\}
\\
+\frac{1}{\hat x_1-\hat x_2}\left\{ \hat x_1 \left[
\text{Li}_2\left(\frac{1}{1-\hat x_1}\right) -
\text{Li}_2\left(\frac{1}{\hat x_1}\right) \right] -\hat x_2 \left[
\text{Li}_2\left(\frac{1}{1-\hat x_2}\right) -
\text{Li}_2\left(\frac{1}{\hat x_2}\right) \right] \right\}
\end{multline} 
\begin{multline}
g(r,r_1,r_2) = \frac{r_1-r_2}{2 \left( y_1-y_2\right)} \left\{ \log
r_1 \left[ \log \left( 1- \frac{r_1}{y_1}\right) - \log \left( 1-
\frac{r_1}{y_2}\right) \right] \right. \\ 
\left.
-\log r_2 \left[ \log
\left( 1- \frac{r_2}{y_1}\right) - \log \left( 1-
\frac{r_2}{y_2}\right) \right] 
+ \text{Li}_2\left(\frac{r_1}{y_1}\right)-\text{Li}_2
\left(\frac{r_2}{y_1}\right)-\text{Li}_2\left(\frac{r_1}
{y_2}\right)+\text{Li}_2\left(\frac{r_2}{y_2}\right) \right\} \\ 
+\frac{1}{2(x_1-x_2)}\left[ \text{Li}_2\left(\frac{1}{1-x_1}\right) -
\text{Li}_2\left(\frac{1}{1-x_2}\right)
-\text{Li}_2\left(\frac{1}{x_1}\right) +
\text{Li}_2\left(\frac{1}{x_2}\right)\right]
\end{multline}
where
\globallabel{eq:roots}
\begin{align}
x_{1,2} &= \frac{1}{2r} \left[ r-r_1+r_2 \mp\sqrt{(r-r_1+r_2)^2-4rr_2}\right] & 
\hat x_i &= x_i|_{r=1} \mytag \\
y_{1,2} &=r_2+(r_1-r_2)x_{1,2} & \hat y_i  &= y_i|_{r=1}. \mytag
\end{align}
In the physical region, the functions $f_{\gamma H}$, $f_{ZH}$ and $f_{WW}$
do not develop imaginary parts.

In order to study the case of chargino masses larger than $M_Z$, or
$r_1,r_2\gg 1$, it is useful to switch to the variables $R =
\sqrt{r_1r_2}$, $\rho=r_1/r_2$ and expand in $1/R$. We then get
\globallabel{eq:fexpansion}
\begin{gather}
f_{\gamma H}(R) = \frac{2+\log R}{2R} + \ord{\frac{\log R}{R^2}}
\mytag \\ 
f_{ZH}(r,r_1,r_2) = a_{ZH}(\rho)\frac{\log R}{R}
+\frac{b_{ZH}(r,\rho)}{R}+ \ord{\frac{\log R}{R^2}} \mytag \\
f_{WW}(r_1,r_2) = a_{WW}(\rho)\frac{\log R}{R}
+\frac{b_{WW}(\rho)}{R}+ \ord{\frac{\log R}{R^2}} \mytag\end{gather}
where \globallabel{eq:fexpansioncoe}

\begin{align}
a_{ZH}(\rho) &= \frac{\sqrt{\rho}}{2(\rho-1)}\;\log \rho \mytag \\
b_{ZH}(r,\rho) &= \frac{\sqrt{\rho}}{2(\rho-1)}\; \left[ \frac{r\log
r\log \rho}{1-r}-\text{Li}_2(1-\rho)+\text{Li}_2(1-1/\rho)\right]
\mytag \\ 
a_{WW}(\rho) &= \frac{\sqrt{\rho}}{(\rho-1)^2}\;\left(
\rho-1-\log \rho \right) \mytag \\ 
b_{WW}(\rho) &=
\frac{\sqrt{\rho}}{(\rho-1)^2} \;\left[ \rho-1+\frac{(\rho+1)}{2}\log
\rho +\text{Li}_2(1-\rho)- \text{Li}_2(1-1/\rho)\right] . \mytag
\end{align}

From these expansions, the expressions for the 
functions $F_{\gamma H}$, $F_{ZH}$,
$F^{(1)}_{WW}$, $F^{(2)}_{WW}$ in \eqs{expansion} follow:
\globallabel{eq:Fexpansion}
\begin{align}
F_{\gamma H}(\rho, R) &= -\frac{1}{2}\log R -1 +\frac{(\rho+1)\log
\rho} {4(\rho-1)} +\ord{\frac{\log R}{R}} \mytag \\ 
F_{ZH}(r,\rho,R)
&= A_{ZH}(\rho)\log R +B_{ZH}(r,\rho) + \ord{\frac{\log R}{R}} \mytag
\\ 
F^{(1)}_{WW}(\rho,R)& = A^{(1)}_{WW}(\rho)\log R
+B^{(1)}_{WW}(\rho)+\ord{\frac{\log R}{R}} \mytag \\
F^{(2)}_{WW}(\rho,R) &= A^{(2)}_{WW}(\rho)\log R +B^{(2)}_{WW}(\rho)
+\ord{\frac{\log R}{R}} , \mytag
\end{align}
where \globallabel{eq:Fexpansioncoe}
\begin{align}
A_{ZH}(\rho) &= \frac{(\rho-1) (2- \rho) - \rho \log
\rho}{4(\rho-1)^2} +\frac{s_W^2}{2} \mytag \\
\begin{split}
B_{ZH}(r,\rho) &= \frac{1}{{4(r-1)(\rho-1)^2}}\left\{
\left( 2-2r+r\log r \right) \left( \rho -1 \right)
\left[ \rho -2 -2 s_W^2 (\rho -1)\right] \right. \\ &
+(r-1) (\rho -1)
\left[ \frac{\rho}{2}+1-s_W^2 (\rho +1)\right] \log \rho 
+ r\rho \log r   \log \rho \\ &
\left.  +(r-1)
\rho\left[ \text{Li}_2(1-\rho)-\text{Li}_2(1-1/\rho)\right] \right\}
\end{split} \mytag \\
A^{(1)}_{WW}(\rho) &= \rho \; \frac{-\rho^2+1+2\rho\log
\rho}{8(\rho-1)^3}\mytag \\ 
B^{(1)}_{WW}(\rho) &= \rho\;
\frac{-4\rho(\rho-1)+(\rho^2-4\rho-1)\log \rho-4\rho
\left[\text{Li}_2(1-\rho)-\text{Li}_2(1-1/\rho)\right]}{16(\rho-1)^3}\mytag \\
A^{(2)}_{WW}(\rho) &= \rho\;\frac{(\rho-7)(\rho-1)+2(\rho+2)\log
\rho}{8(\rho-1)^3} \mytag \\ 
B^{(2)}_{WW}(\rho) &=
\rho\;\frac{4(\rho-4)(\rho-1)-(\rho^2+4\rho+7)\log \rho
-4(\rho+2)\left[ \text{Li}_2(1-\rho)-\text{Li}_2(1-1/\rho)
\right]}{16(\rho-1)^3}
. \mytag
\end{align}

\end{document}